\documentclass[apjl]{emulateapj}

\usepackage{graphicx,amsmath,amsfonts,amssymb,hyperref}

\begin{document}

\title{The discovery of seven extremely low surface brightness galaxies in the
field of the nearby spiral galaxy M101}

\author{Allison Merritt\altaffilmark{1}, Pieter van Dokkum\altaffilmark{1}, \& Roberto Abraham\altaffilmark{2}}

\altaffiltext{1}
{Department of Astronomy, Yale University, 260 Whitney Avenue, New Haven, CT, USA 06511}
\altaffiltext{2}
{Department of Astronomy and Astrophysics, University of Toronto,
50 St. George Street, Toronto, ON, Canada M5S~3H4}

\begin{abstract}
Dwarf satellite galaxies are a key probe of dark matter and of galaxy formation on small
scales and of the dark matter halo masses of their central galaxies. They have very low
surface brightness, which makes it difficult to identify and study them outside of the
Local Group. We used a low surface brightness-optimized telescope, the Dragonfly Telephoto
Array, to search for dwarf galaxies in the field of the massive spiral galaxy M101. We
identify seven large, low surface brightness objects in this field, with effective radii
of \(10 - 30\) arcseconds and central surface brightnesses of \(\mu_{g} \sim 25.5 - 27.5\)
mag arcsec\(^{-2}\). Given their large apparent sizes and low surface brightnesses, these
objects would likely be missed by standard galaxy searches in deep fields. Assuming the
galaxies are dwarf satellites of M101, their absolute magnitudes are in the range
\(-11.6 \lesssim M_{V} \lesssim -9.3\) and their effective radii are \(350\) pc \(-\)
\(1.3\) kpc. Their radial surface brightness profiles are well fit by Sersic profiles with
a very low Sersic index (\(n \sim 0.3 - 0.7\)). The properties of the sample are similar
to those of well-studied dwarf galaxies in the Local Group, such as Sextans I and Phoenix.
Distance measurements are required to determine whether these galaxies are in fact
associated with M101 or are in its foreground or background.
\end{abstract}


\section{Introduction} 
In recent years, the number of known dwarf galaxies residing within the Local Group has
increased dramatically (see McConnachie 2012, and references therein). The Milky
Way and Andromeda (M31) galaxies are each host to dozens of faint satellites
ranging in central surface brightness from \(20-30\) mag arcsec\(^{-2}\), most of which
have been uncovered in star count surveys
(e.g., {Ibata} {et~al.} 2007; {Belokurov} {et~al.} 2010; {Richardson} {et~al.} 2011;
{Martin} {et~al.} 2013, and several others). 

Through measurements of their kinematics, Local Group satellites provide constraints on
the masses of the dark matter halos of the Milky Way and M31
(e.g., {Battaglia} {et~al.} 2005; {Watkins}, {Evans}, \& {An} 2010). They also serve as
testing sites for theories of cosmology and galaxy evolution on small scales. Comparisons
of observed satellite abundances and internal structure with predictions from
\(\Lambda\)CDM, for example, have led to the now familiar ``missing satellite'' problem
({Kauffmann}, {White}, \&  {Guiderdoni} 1993; {Klypin} {et~al.} 1999;
{Moore} {et~al.} 1999) and the ``too big to fail'' problem
({Boylan-Kolchin}, {Bullock}, \&  {Kaplinghat} 2012), respectively. To more robustly
determine the magnitude of the challenges faced by \(\Lambda\)CDM, however, we will need
to expand the sample size beyond the Local Group.

Most dwarf galaxies have extremely low surface brightness. If the known Milky Way
satellites were located at \(5\) Mpc their median integrated apparent magnitude would be
\(m_{V} \sim 21.8\), but their median central surface brightness would be
\(\mu_{0,V} \sim 26.1\), too faint to be detected in most integrated light surveys.
Studies based on star counts are able to reach surface brightnesses of \(30\) mag
arcsec\(^{-2}\) or fainter \(-\) but only for very nearby galaxies, as the brightness of
stars, and thus the number of detectable tracer stars in a distant source, decreases with
the square of the distance.\footnote{As an example, dwarfs with \(M_{V} \sim -8\) can be
detected out to \(\sim 1\) Mpc with SDSS ({Koposov} {et~al.} 2008).} By contrast,
integrated light surface brightness is conserved with distance, and the development of
integrated light techniques sensitive enough to allow dwarfs to be detected beyond the
Local Group could expand the number of known dwarfs by orders of magnitude.

Several dwarf galaxies have already been identified by their low surface brightness
appearance in integrated light (e.g., {Karachentsev} {et~al.} 2014). Furthermore, a number
of technological advances that promise to make imaging of very low surface brightness
galaxies routine have recently been developed ({Abraham} \& {van Dokkum} 2014). It is
within this context that we describe results from a search for faint dwarf galaxies around
the nearby spiral galaxy M101. We find seven previously unknown low surface brightness
(LSB) galaxies, and we assess the likelihood that they are members of the M101 group.

\section{Data Collection and Reduction}

\begin{deluxetable*}{cccccccccccc}
\tabletypesize{\footnotesize}
\tablecolumns{12}
\tablewidth{0pt}
\tablecaption{Observed Properties of the Sample \label{paramstable}}
\tablehead{
\colhead{ID} & \colhead{$\alpha$} & \colhead{$\delta$} & \colhead{$g$\,\tablenotemark{a}} &
\colhead{$\mu_{0,g}$\,\tablenotemark{b}} & \colhead{$\mu_{e,g}$\,\tablenotemark{c}} &
\colhead{$g-r$} & \colhead{$r_{e}$\,\tablenotemark{d}} &
\colhead{$n$\,\tablenotemark{e}} & \colhead{$b/a$\,\tablenotemark{f}} \\
\colhead{} & \colhead{(J2000)} & \colhead{(J2000)} & \colhead{} &
\colhead{} & \colhead{} & \colhead{} & \colhead{} & \colhead{} & \colhead{}
}
\startdata
DF\_1 & 14 03 45.0 & +53 56 40 & 18.9 $\pm$ 0.1 & 25.6 $\pm$ 0.1 & 26.6 $\pm$ 0.1 & 0.5 $\pm$ 0.2 & 14 $\pm$ 1 & 0.6 $\pm$ 0.1 & 0.6 $\pm$ 0.1 \\ 
DF\_2 & 14 08 37.5 & +54 19 31 & 19.4 $\pm$ 0.2 & 25.8 $\pm$ 0.3 & 26.9 $\pm$ 0.2 & 0.5 $\pm$ 0.2 & 10 $\pm$ 1 & 0.7 $\pm$ 0.2 & 0.9 $\pm$ 0.1 \\ 
DF\_3 & 14 03 05.7 & +53 36 56 & 17.9 $\pm$ 0.2 & 26.4 $\pm$ 0.2 & 27.4 $\pm$ 0.2 & 0.6 $\pm$ 0.2 & 30 $\pm$ 3 & 0.6 $\pm$ 0.1 & 0.7 $\pm$ 0.1 \\ 
DF\_4 & 14 07 33.4 & +54 42 36 & 18.8 $\pm$ 0.3 & 26.8 $\pm$ 0.4 & 28.0 $\pm$ 0.2 & 0.6 $\pm$ 0.4 & 28 $\pm$ 7 & 0.7 $\pm$ 0.3 & 0.6 $\pm$ 0.1 \\ 
DF\_5 & 14 04 28.1 & +55 37 00 & 18.0 $\pm$ 0.2 & 27.4 $\pm$ 0.3 & 28.0 $\pm$ 0.2 & 0.5 $\pm$ 0.4 & 38 $\pm$ 7 & 0.4 $\pm$ 0.2 & 0.8 $\pm$ 0.1 \\ 
DF\_6 & 14 08 19.0 & +55 11 24 & 20.1 $\pm$ 0.4 & 27.5 $\pm$ 1.1 & 27.8 $\pm$ 0.4 & 0.4 $\pm$ 0.5 & 22 $\pm$ 8 & 0.3 $\pm$ 0.8 & 0.3 $\pm$ 0.1 \\ 
DF\_7 & 14 05 48.3 & +55 07 58 & 20.4 $\pm$ 0.6 & 27.7 $\pm$ 1.6 & 28.7 $\pm$ 0.6 & 0.9 $\pm$ 0.8 & 20 $\pm$ 9 & 0.6 $\pm$ 1.0 & 0.5 $\pm$ 0.2 \\ 
\enddata
\tablenotetext{a}{Integrated apparent magnitude.}
\tablenotetext{b}{Central surface brightness, in mag arcsec$^{-2}$.}
\tablenotetext{c}{Effective surface brightness, in mag arcsec$^{-2}$.}
\tablenotetext{d}{Effective radius, in arcseconds.}
\tablenotetext{e}{Sersic index.}
\tablenotetext{f}{Axis ratio.}
\tablecomments{Structural parameters were computed using GALFIT from a stack of the \(g\)- and \(r\)-band images.}
\end{deluxetable*}

Imaging faint galaxies in integrated light requires a telescope capable of detecting very
low surface brightness emission. Our observations were taken with the Dragonfly Telephoto
Array, a new robotic refracting telescope designed specifically for this purpose
({Abraham} \& {van Dokkum} 2014). The telescope is comprised of an array of eight 400 mm
\(f/2.8\) Canon IS II telephoto lenses which, when operating together, are equivalent to
an \(f/1\) optical system.\footnote{This is crucial for low surface brightness imaging, as
the counts per unit area on the detector decrease inversely with the square of the focal
ratio \(f\).} Nano-fabricated coatings on the optical elements of these lenses suppress
internally scattered light -- typically a significant obstacle to low surface brightness
imaging ({Slater} {et~al.} 2009) -- by an order of magnitude. The Dragonfly field
of view covers \(2.6^{\circ} \times 1.9^{\circ}\) in a single frame, and
\(3.3^{\circ} \times 2.8^{\circ}\) once dithered frames have been combined. At an assumed
distance to M101 of 7 Mpc ({Shappee} \& {Stanek} 2011; {Lee} \& {Jang} 2012), this
corresponds to \(\sim 403 \times 342\) kpc. For reference, the virial radius of M101 is
\(\sim 260\) kpc.\footnote{Using a stellar mass of
\(M_{*} \sim 5.3\times 10^{10} M_{\odot}\) ({van Dokkum} {et~al.} 2014) in combination
with the stellar mass - halo mass relation given by Moster {et~al.} (2010), we estimate a
halo mass of \(M_{\rm h}\sim 2\times 10^{12}M_{\odot}\), which corresponds to a virial
radius of \(\sim 260\) kpc.}

Details of data acquisition and reduction are provided in {van Dokkum} {et~al.} (2014);
here we give only a brief description. The data were taken during May and June of 2013 for
a total of 35 hours. Calibration frames were taken on each night and applied to individual
images, along with an illumination correction and an additional correction for the sky
gradient. Images on a given night and across different nights were combined using optimal
weighting. The final data product is comprised of reduced and star-subtracted frames in
the \textit{g}-band and \textit{r}-band. The limiting surface brightness in the final
reduced images is \(\mu_{g} \sim 29.5\) mag arcsec\(^{-2}\) and \(\mu_{r} \sim 29.8\) mag
arcsec\(^{-2}\) on scales of \(\sim 10\) arcsec.

\begin{figure*}[!t]
\begin{center}
\includegraphics[scale=0.8]{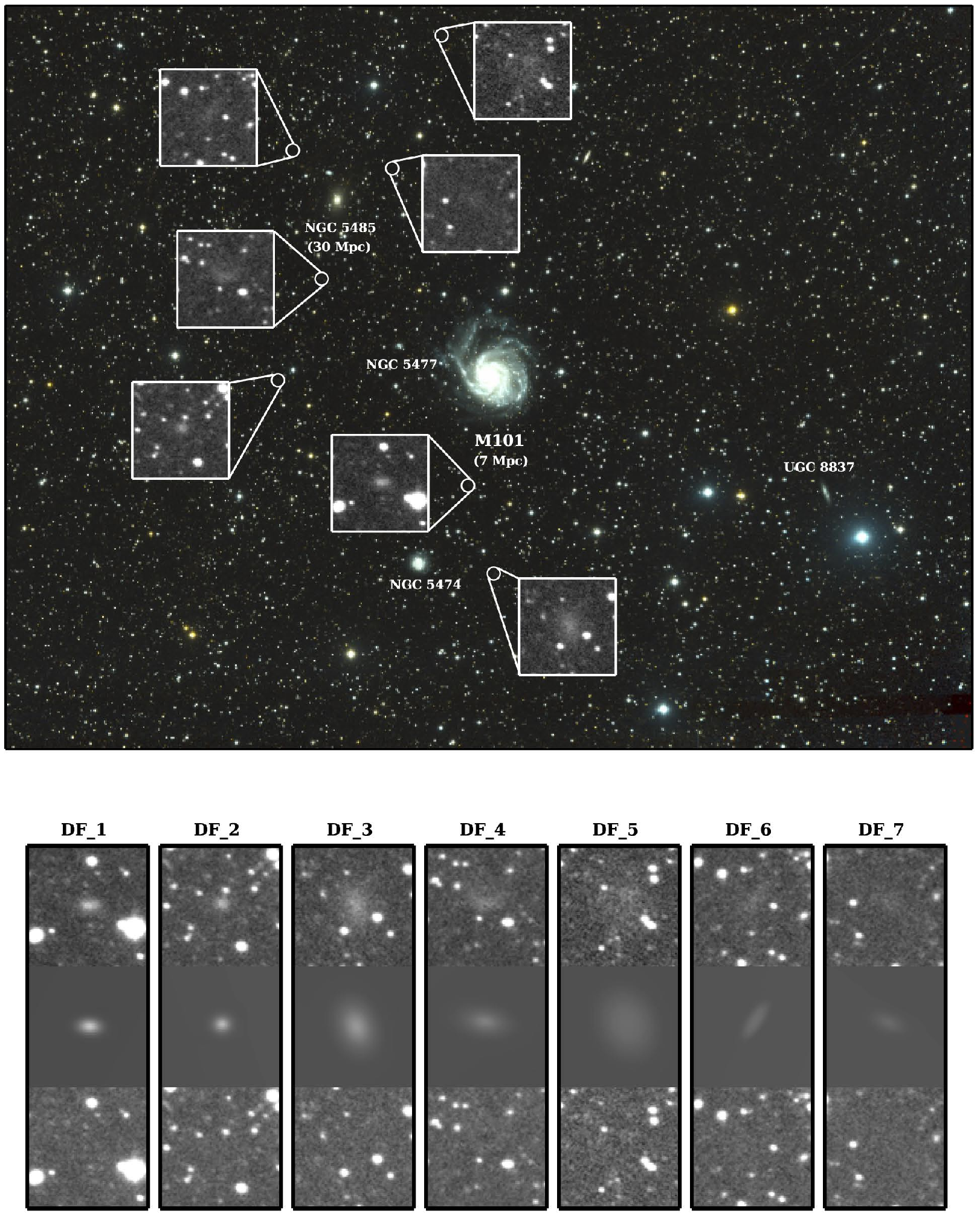}
\caption{\textbf{Top:} The full \(3.3^{\circ} \times 2.8^{\circ}\) Dragonfly field of
	view, centered on M101. The zoomed cutouts highlight the position of each of the seven
	newly discovered LSBs. North is up, and East is to the left. The five additional
	members of the M101 group that fall within our field of view (NGC 5474, NGC 5477, UGC
	8837, UGC 8882) are labeled, as are the HI cloud GBT 1355+5439 and the background
	galaxy NGC 5485. \textbf{Bottom:} From left to right we show each of the seven LSBs.
	In each panel we show the central \(100 \times 100\) pixels of the \(g\)-band cutouts,
	\(r\)-band cutouts, \(g\)-band GALFIT fits, and associated \(g\)-band residuals.
	\label{fovandcutouts}}
\end{center}
\end{figure*}

\section{A Search for Low Surface Brightness Objects}

Six of the seven LSB galaxies presented here were initially discovered in the vicinity
of M101 by visual inspection. Motivated by this, we developed a simple algorithm to allow
semi-automatic detections of these and similar objects in the field (see {Vollmer}
{et~al.} (2013) for an example of more sophisticated methods). Our algorithm recovered the
six visually-identified objects, and detected one additional galaxy (DF\_3). The method is
described below; it utilizes the reduced \textit{g}-band and \textit{r}-band images as
well as their star-subtracted counterparts (see {van Dokkum} {et~al.} 2014).

In order to detect all objects in each image, SExtractor ({Bertin} \& {Arnouts} 1996) was
run three separate times, using varied convolution filters and detection thresholds, and
the three SExtractor output source catalogs were subsequently combined for each frame. The
first requirement for an LSB detection was that it appeared in both \(g\) and \(r\)
frames, and we therefore combined catalogs for the two filters (for reduced and
star-subtracted frames separately), excluding any detections that were only found in a
single filter. Objects were matched based on their positions and relative sizes. Extended
LSB objects were not removed from the star-subtracted frames along with the stars, so we
further required that objects were detected there as well as in the original (i.e.,
pre-star subtraction) frames. Finally, we imposed conservative constraints on the size
(\(5 \leq R \leq 50\) pixels), median count level (\(\leq 0.03\) counts) and scatter
(\(\leq 0.008\) counts) for detections in order to optimize the search for LSB objects. 

This selection reduced the number of detections that require visual inspection from
\(\sim 108,098\) sources in the $g$-band catalog to only \(529\). Six of these
corresponded to the original sample that was found by eye, and one additional LSB object
was identified. Of the remaining \(522\), \(28\)\% were either wings of bright stars that
were not fully subtracted, or in close proximity to stars or galaxies; \(19\)\% were false
positives caused by closely spaced faint stars; and the final \(53\)\% were deemed to be
noise fluctuations. Deeper observations may reveal that some candidates in this latter
group are galaxies as well. An important caveat here is that our search is insensitive
to the smallest galaxies; the FWHM of stars in our images is \(\sim 6.5\) arcseconds,
corresponding to \(\sim 200\) pc at the distance of M101. 

To quantify how efficiently our algorithm detects LSB objects, we simulated LSBs with
a range of central surface brightness and size (\(30 \leq \mu_{0} \leq 23\) mag
arcsec\(^{-2}\), \(6 \leq r_{e} \leq 50\) arcsec) and placed them at random locations in
our data. The algorithm was then run on these images to determine how well the simulated
LSBs were recovered. We found that \(\sim 70\)\% of the simulated LSBs with properties
similar to those of our observed sample were detected, but detection rates drops for
other combinations of size and surface brightness. We therefore interpret our seven
detections as a lower limit for the number of LSBs in the field of view.

\begin{figure*}[!t]
\begin{center}
\includegraphics[scale=0.8]{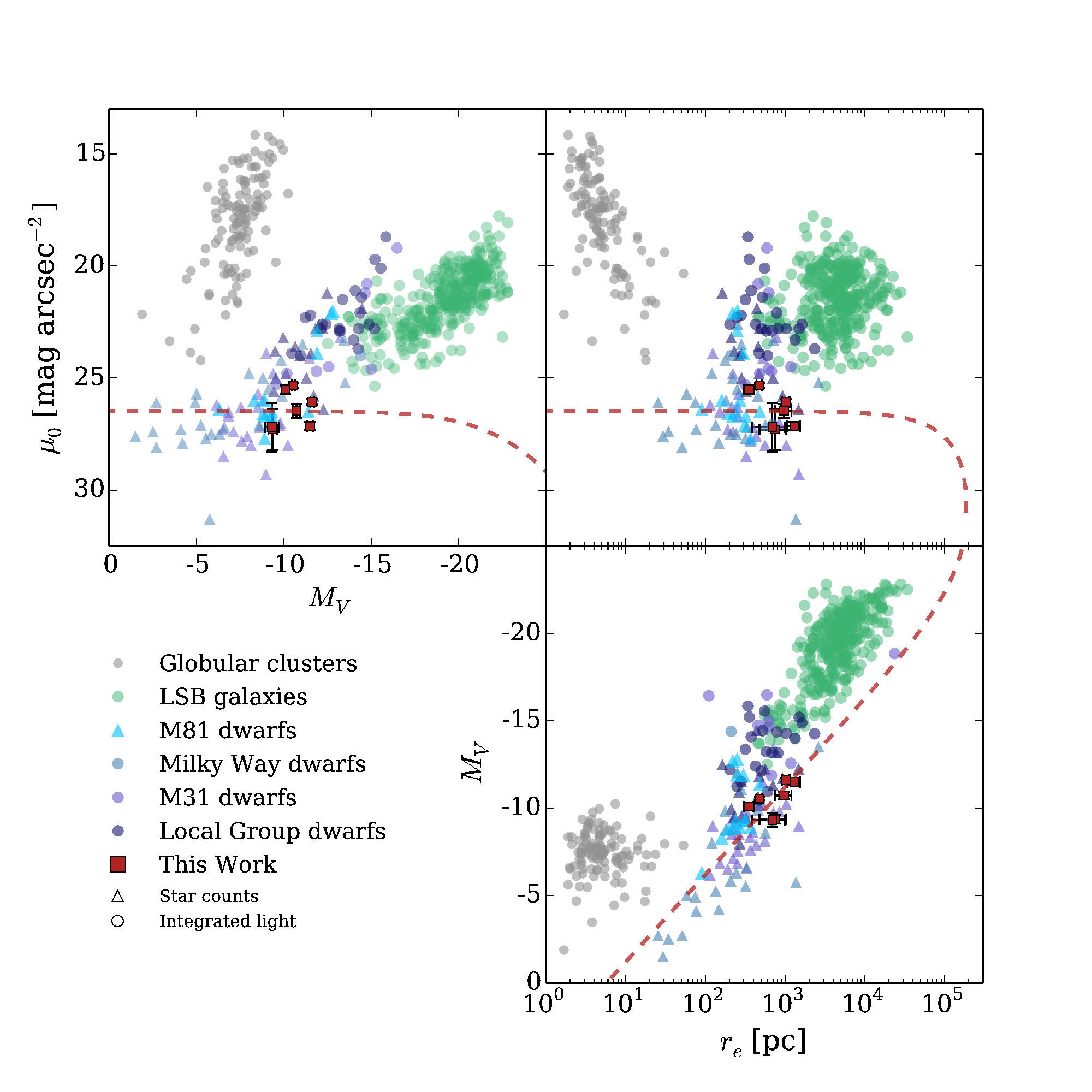}
\caption{Comparison of the range in surface brightness, absolute \(V\) magnitude and
	effective radius of our LSB sample with that of the Local Group dwarfs (as compiled
	by McConnachie 2012, see also references therein), nearby field LSBs
	({Impey} {et~al.} 1996), M81 dwarfs ({Chiboucas} {et~al.} 2013) and galactic globular
	clusters (Harris 1996). The \(V\) magnitudes of the M81 dwarfs were converted from
	\(r\)-band using the conversions of {Fukugita} {et~al.} 1996 and the
	\(\langle B-V \rangle\) color of the Local Group dwarfs, and we assume
	\(\langle B-V \rangle \sim 0.43\) ({Romanishin} {et~al.} 1983) for the
	field LSBs. We use a distance of \(7\) Mpc - the red dashed line indicates where
	the median properties of our sample would fall for distances from the edge of the
	Milky Way halo out to \(z \sim 2\) (including the effects of cosmological dimming).
\label{sbfigs}}
\end{center}
\end{figure*}

\begin{figure*}[!t]
\begin{center}
\includegraphics[scale=0.7]{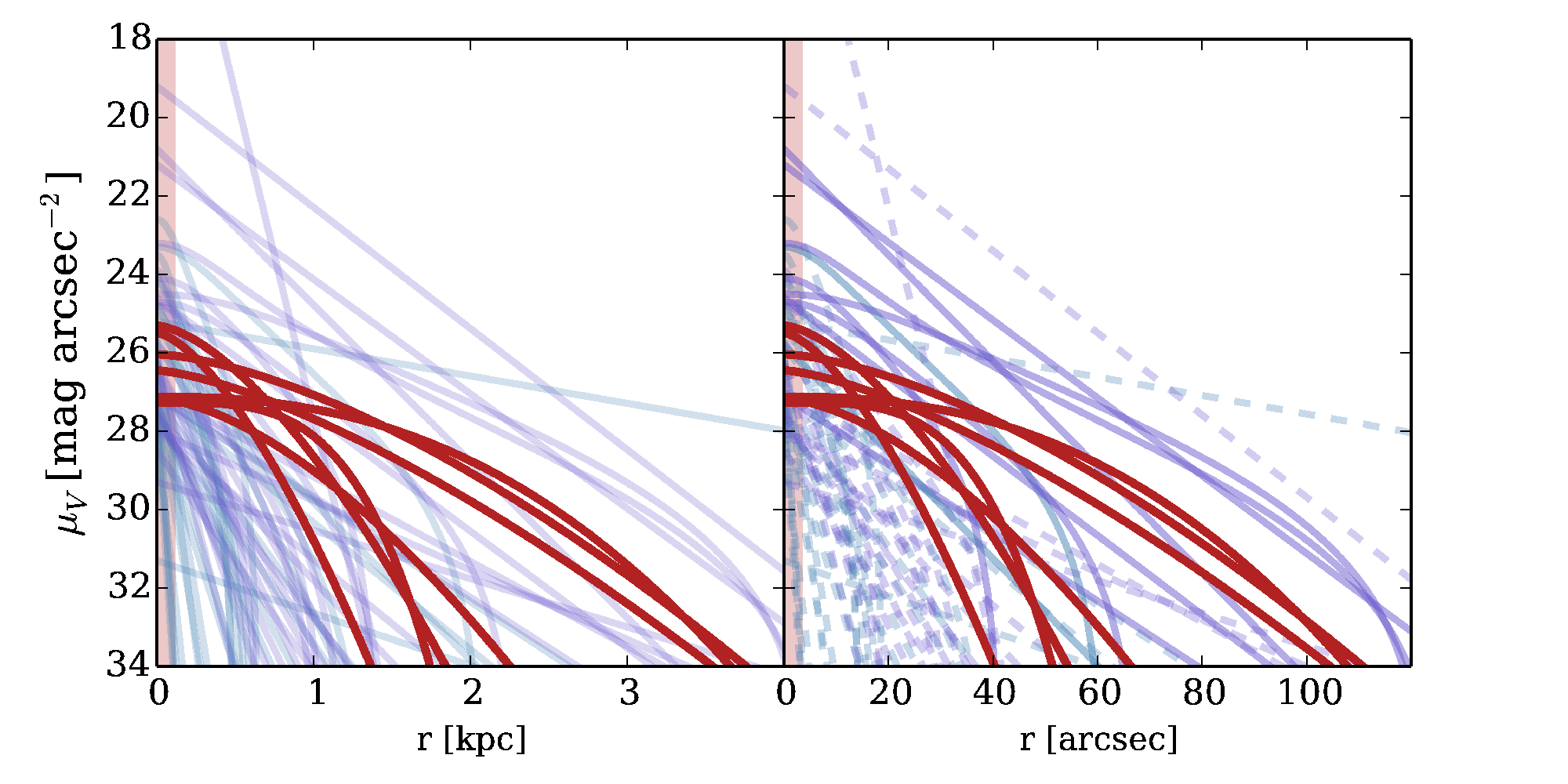}
\caption{Radial profiles of the seven LSBs (red), constructed from the Sersic parameters
	measured by GALFIT. Blue and purple lines represent the profiles of dwarf satellites
	of the Milky Way and M31, respectively, and the shaded region corresponds to FWHM/2
	for our data. \textbf{Left}: the physical properties of the Local Group satellites as
	well as those of our sample for a distance of 7 Mpc. \textbf{Right}: the
	\textit{observed} properties of our sample and the implied observed properties if
	the Local Group satellites were at 7 Mpc. Dashed lines indicate redshifted Local Group
	dwarfs which, when modeled with GALFIT and placed into our data, were not detected by
	our algorithm.
\label{radialfig}}
\end{center}
\end{figure*}

\section{Structure and Brightness}
We used GALFIT (Peng {et~al.} 2002) to determine the structure, luminosity, and
surface brightness of the galaxies. We chose a region of \(200 \times 200\) pixels around
each LSB and masked out any nearby stars. We first simultaneously fit the galaxy with a
{Sersic} (1968) profile and any overlapping stars with delta functions. Both the Sersic
models and delta functions were convolved with the Dragonfly PSF. A model was produced for
the stars, and they were subsequently removed from the foreground of the images. Next, we
stacked the \textit{g}-band and \textit{r}-band star-subtracted cutouts, and ran GALFIT a
second time to measure the structure and orientation of the galaxies at higher S/N.
Finally, we fit the luminosity and surface brightness of the galaxies in \textit{g}-band
and \textit{r}-band individually, holding all other parameters fixed at their previously
determined values. Figure \ref{fovandcutouts} shows the cutouts, best fit model, and
associated residuals for DF\_1 - DF\_7, and Table \ref{paramstable} contains the values
for each parameter. 

We find that the measured surface brightnesses of the LSBs are low, ranging in central
surface brightness from \(\sim 25.5 - 27.5\) mag arcsec\(^{-2}\) in \textit{g}-band with
corresponding surface brightnesses of \(26.5 - 28.5\) mag arcsec\(^{-2}\) measured at the
effective radius. Effective radii range from \(10 - 30\) arcseconds. 

The dominant source of error in our measurements of these parameters is the low
signal-to-noise ratio in the fitting regions. To quantify this, we placed our best fit
model of each LSB in 100 relatively empty, random locations in the M101 field. Each time,
we re-measured every parameter, applying the same steps that were used for the original
sample. The scatter in the values for each parameter represents the uncertainty in the fit
due to noise and systematic errors such as background estimation. The errors are listed in
Table \ref{paramstable}.

We note that these galaxies are not detected in the Sloan Digital Sky Survey ({Abazajian}
{et~al.} 2009), although the central regions of the brightest few are visible in SDSS
images.

\section{A new population of faint dwarfs in the M101 group?}
Given their location relative to M101 (all seven lie within its projected virial radius),
we consider the possibility that these seven new LSB galaxies are dwarf satellite
galaxies. In the absence of available distance measurements for the LSBs, we use
comparisons of their physical properties (computed at a distance of 7 Mpc) with those of
known Local Group dwarf satellite galaxies. If the LSBs are satellites of M101, we may
expect that the properties of the two populations will be consistent with one another.

In Figure \ref{sbfigs}, we plot central surface brightness as a function of effective
radius and absolute magnitude for our sample and the Local Group dwarfs.\footnote{We
converted our magnitudes from \textit{g} to \(V\) using the transformations given in
{Fukugita} {et~al.} (1996).} The LSBs presented here have a median central surface
brightness of \(\mu_{0,V}\sim 26.5\) mag arcsec\(^{-2}\), which is very close to the
median values for the MW and M31 dwarf satellites (\(26.6\) and \(26.3\) mag
arcsec\(^{-2}\), respectively). The median absolute magnitude \(M_{V} \sim -10.5\) is also
typical of the Local Group dwarfs. Furthermore, we find that the integrated colors are
similar: the median color of the LSB sample is \(\langle g-r \rangle \sim 0.5\), or
\(\langle B-V \rangle \sim 0.7\), whereas the brighter Local Group dwarfs (which have
\(M_{V}\) comparable to our sample) have a median color of
\(\langle B-V \rangle \sim 0.63\) ({Mateo} 1998, and references therein). The sizes of the
LSBs span a range that is very similar to that of the M31 satellites, but are consistent
with only the largest of the known Milky Way satellites. As noted previously, we cannot
detect the smallest satellites due to the \(6.5\) arcsecond resolution of Dragonfly.

Finally, we compare the internal structure of the galaxies. Every galaxy in our sample has
a Sersic index of \(n < 1\); the median value is \(n \sim 0.6\). Dwarf satellites in the
Local Group are typically fit with either an \(n = 1\) profile (e.g.,
{Ibata} {et~al.} 2007; {Richardson} {et~al.} 2011) or with a King profile (e.g., 
{Irwin} \& {Hatzidimitriou} 1995; {de Jong} {et~al.} 2008) \(-\) the latter are similar to
\(n < 1\) Sersic profiles, as both have a deficit of light at the center and in
the outskirts relative to an exponential profile. This is borne out in Figure
\ref{radialfig}, where we show the radial surface brightness profiles of our LSBs and of
the Local Group dwarfs (using structure as reported by {McConnachie} 2012). In the left
panel, we show the profiles of our LSBs (at 7 Mpc) alongside the profiles of the Local
Group dwarfs, and in the right panel we show the observed profiles of the LSBs with
redshifted profiles of the Local Group dwarfs. The light profiles of the LSBs and the
Local group dwarfs encompass a range of properties, but are consistent with one another.

\section{Discussion}
In this \textit{Letter} we have presented the discovery of seven LSB galaxies in the field
of the nearby spiral galaxy M101. The galaxies in our sample range from \(25.5-27.5\) mag
arcsec\(^{-2}\) in central surface brightness and have Sersic indices \(n < 1\). As shown
in Figures \ref{sbfigs} and \ref{radialfig}, the properties of the LSBs are similar to
those of dwarf satellites in the Local Group, but we stress that without distance
measurements for these galaxies other interpretations are possible. Here we provide a
discussion of the implications of associating the LSBs with M101 and also explore some
additional scenarios.

\begin{figure}[!t]
\begin{center}
\includegraphics[scale=0.55]{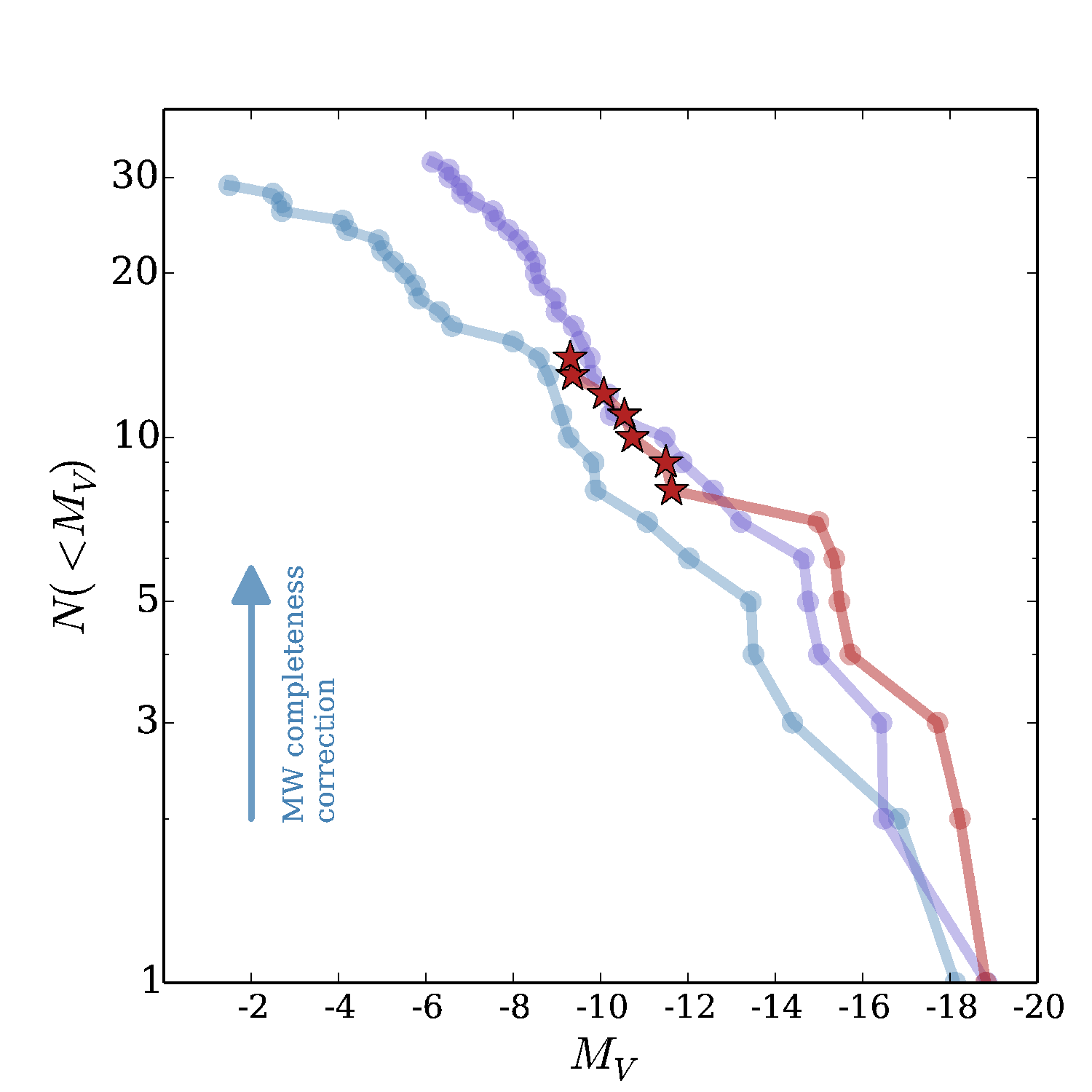}
\caption{The observed cumulative luminosity function of the M101 group (red), including
	the LSBs presented here. We compare this to the observed CLFs of the Milky Way (blue)
	and M31 (purple), using data from McConnachie (2012). The arrow represents the
	factor of \(\sim 3\) completeness correction for the Milky Way CLF.
\label{clffig}}
\end{center}
\end{figure}

To date, the M101 group is known to consist of seven relatively bright companions
(\(-19 < M_{V} < -14\); {Giuricin} {et~al.} (2000); {Karachentsev} {et~al.} 2014), five of
which fall within our field of view. Additionally, {Mihos} {et~al.} (2012) discovered two
HI clouds in the vicinity of M101 - one, GBT 1355+5439, lies in our field, but we do not
detect any signal above the limiting surface brightness (see Section 2) at that location.
In Figure \ref{clffig} we show the cumulative luminosity function (CLF) for the M101 group
with the LSBs included, along with the observed CLFs of the Milky Way and M31 for
comparison. We note that for \(M_{V} \lesssim -9\), the M101 group CLF is remarkably
similar to that of M31. Another point of interest is the apparent arrangement of the LSB
galaxies - all seven were discovered to the east of M101. Particularly in the context of
the lack of observed tidal streams or stellar halo down to \(\gtrsim 30\) mag
arcsec\(^{-2}\) (Mihos {et~al.} 2013; {van Dokkum} {et~al.} 2014), this may indicate that
the galaxies are part of an infalling low mass group (e.g., {Tully} {et~al.} 2006;
{Wetzel} {et~al.} 2013).

We also consider the possibility that the galaxies in this sample are not associated with
M101. Measured central surface brightnesses for field LSBs out to \(z \sim 0.1\) (green
points, {Impey} {et~al.} 1996) are shown in Figure \ref{sbfigs}, along with absolute
magnitude and effective radius. It is evident from these plots that the LSBs presented
here have lower surface brightness, and implied lower luminosity and smaller sizes (at
\(7\) Mpc) than the majority of that population. The dashed red line in Figure
\ref{sbfigs} shows how these properties change as a function of distance. At large
distances, our sample of LSBs would be considerably fainter than the main LSB population
at fixed \(M_{V}\). The interpretation of this discrepancy is not straightforward,
however, as the limiting surface brightness of the {Impey} {et~al.} (1996) catalog is
\(26\) B mag arcsec\(^{-2}\). We therefore cannot rule out the possibility that a
significant population of large, very low surface brightness field galaxies at
intermediate redshift has gone undetected thus far. However, their \(n < 1\) structure
would be very different from known field LSBs, which often have both an exponential disk
and a bulge component (e.g., {Romanishin} {et~al.} 1983). Several of the galaxies are
projected near the background galaxy NGC 5485 (indicated in Figure \ref{fovandcutouts},
located at a distance of \(30\) Mpc). If these galaxies are satellites of NGC 5485, their
median absolute magnitude and effective radius would be \(M_{V} \sim -13.7\) and
\(r_{e} \sim 3\) kpc. While it is plausible that a subset of the LSBs belong to this
group, their \(n < 1\) profiles make this scenario unlikely. It is also possible that the
LSBs reside within the halo of the Milky way \(-\) if this is the case, they would have
median sizes of \(\sim 16\) pc and luminosities of \(M_{V}\sim -2.2\). 

Given the faint, diffuse nature of the LSBs, we also assess the likelihood that we are
observing galactic cirrus, planetary nebulae, or globular clusters. Optical studies have
identified galactic cirrus on scales from degrees down to \(\sim 10\) arcseconds
({Guhathakurta} \& {Tyson} 1989). This range encompasses the sizes of the LSBs; however,
the morphologies of the LSBs are inconsistent with the wispy and stratified nature of
cirrus clouds, with the possible exception of DF\_4. Planetary nebulae (PNe) have apparent
magnitudes that are consistent with our LSB sample ({Mal'Kov} 1997); however, PNe are
extremely blue in \(g-r\) due to the presence of [O{\sc III}]\(\lambda 4959,5007\) in the
\(g\)-band (e.g., {Kniazev} {et~al.} 2014). We include a sample of globular clusters (grey
points, Harris 1996) in Figure \ref{sbfigs}, and note that their properties are
inconsistent with both the LSBs and the Local Group dwarfs.

The characterization of these seven new LSBs relies heavily upon determining the distance
to each individual galaxy. Distance measurements for these galaxies will be difficult to
obtain due to their low surface brightnesses, but may be possible with a combination of
spectroscopy and high resolution imaging.

The M101 field was the first in an ongoing photometric survey of nearby galaxies with the
Dragonfly telescope, and we will extend this work to searches for dwarfs around other
galaxies. The discovery of LSB satellite populations around a larger sample of parent
galaxies would not only provide key constraints on cosmology and galaxy evolution on small
scales, but also open up the possibility of measurements of dark matter halo masses for
individual galaxies (e.g., {Zaritsky} \& {White} 1994; {Battaglia} {et~al.} 2005).

\acknowledgements{We thank the referee for helpful comments, and the staff at New Mexico
Skies Observatory for their support throughout this project. Support from NSF grant
AST-1312376 and NSERC is gratefully acknowledged.}


\end{document}